# The grammar of mammalian brain capacity

A.Rodriguez, R.Granger *

*6207 Moore Hall, Dartmouth College, Hanover, NH 03755, United States*

ABSTRACT

Uniquely human abilities may arise from special-purpose brain circuitry, or from concerted general capacity increases due to our outsized brains. We forward a novel hypothesis of the relation between computational capacity and brain size, linking mathematical formalisms of grammars with the allometric increases in cortical-subcortical ratios that arise in large brains. In sum, i) thalamocortical loops compute formal grammars; ii) successive cortical regions describe grammar rewrite rules of increasing size; iii) cortical-subcortical ratios determine the quantity of stacks in single-stack pushdown grammars; iii) quantitative increase of stacks yields grammars with qualitatively increased computational power. We arrive at the specific conjecture that human brain capacity is equivalent to that of indexed grammars – far short of full Turing-computable (recursively enumerable) systems. The work provides a candidate explanatory account of a range of existing human and animal data, addressing longstanding questions of how repeated similar brain algorithms can be successfully applied to apparently dissimilar computational tasks (e.g., perceptual versus cognitive, phonological versus syntactic); and how quantitative increases to brains can confer qualitative changes to their computational repertoire.

## I. Brain growth shows surprisingly few signs of evolutionary pressure

Different animals exhibit different mental and behavioral abilities, but it is not known which abilities arise from specializations in the brain, i.e., circuitry to specifically support or enable particular capacities. Evolutionary constraints on brain construction severely narrow the search for candidate specializations. Although mammalian brain sizes span four orders of magnitude (1), the range of structural variation differentiating those brains is extraordinarily limited.

An animal's brain size can be roughly calculated from its body size (2), but much more telling is the relationship between the sizes of brains and of their constituent parts: the size of almost every component brain circuit can be computed with remarkable accuracy just from the overall size of that brain (1, 3-5), and thus the ratios among brain parts (e.g. cortical to subcortical size ratios) increase in a strictly predictable allometric fashion as overall brain size increases (6, 7) (Figure 1).

These allometric regularities obtain even at the level of individual brain structures (e.g., hippocampus, basal ganglia, cortical areas). There are a few specific exceptions to the well-documented allometric rule (such as the primate olfactory system (8)), clearly demonstrating that at least some brain structure sizes *can* be differentially regulated in evolution, yet despite this capability, it is extremely rare for telencephalic structures ever to diverge from the allometric rule (4, 6, 7, 9). Area 10, the frontal pole, is the most disproportionately expanded structure in the human brain, and has sometimes been argued to be *selected*

**Corresponding author email:* Richard.Granger@gmail.com

for differential expansion, yet the evidence has strongly indicated that area 10 (and the rest of anterior cortex) are nonetheless precisely the size that is predicted allometrically (6, 7, 10, 11).

These findings inexorably lead to the remarkable conclusion that, with few exceptions, brains do not choose which structures to differentially expand or reconfigure (4, 6, 11-15). The same allometric relations recur for all the primary components of the mammalian forebrain (telencephalon), and the same recurring circuit motifs, large and small, are repeated throughout the brain (16, 17). The resulting truly-notable uniformity holds across orders of mammals (along with distinct variants seen within different subgroups such as rodents versus primates (7)).

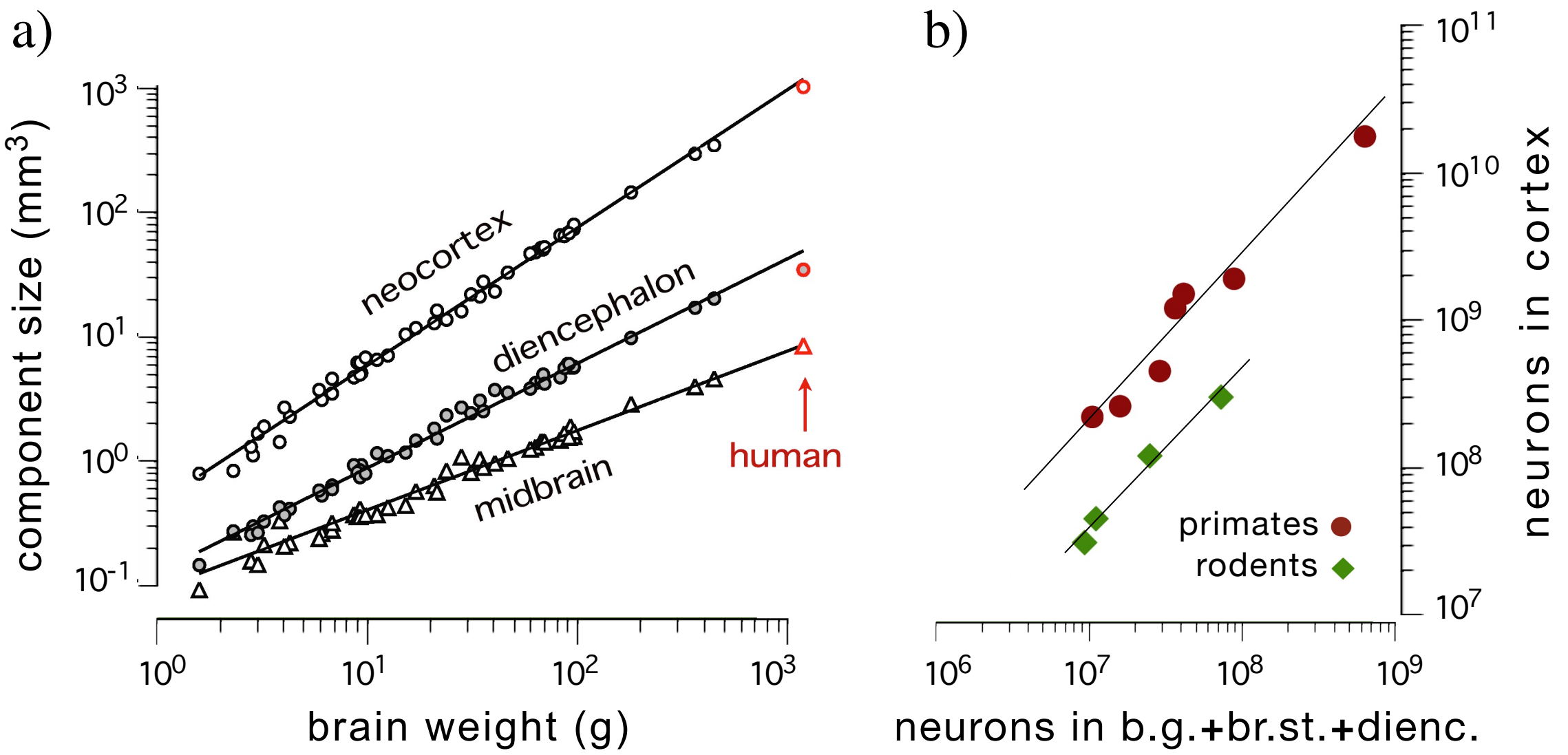

**Figure 1.** Allometric relations among brain parts. (a) Distinct brain components grow at different fixed rates as brain size increases. Shown are the sizes of brains and components in primates. The slope of neocortical increase is greater than 1; neocortex becomes disproportionately enlarged in big brains, in an allometrically highly predictable fashion. Notably, human brains are no exception. (b) Corresponding increases in numbers of neurons accompany cortical size increase, again increasing predictably and allometrically. Shown are cortical neurons vs. the sum of neurons in basal ganglia (b.g.), brain stem (br.st.), and diencephalon (dienc.).

These unexpectedly powerful allometric constraints strongly suggest that novel specialized circuits may not be the explanation for novel abilities in human brains (such as syntax). Rather, this may be an instance in which a quantitative change (increased brain size) results in a qualitative one (uniquely human abilities): simply adding more of the same computational units creates new competencies. Such instances are rare but far from unknown (e.g., in evolution, and in economies of scale). A few specific well-studied computational mechanisms exhibit the ability to yield qualitative changes arising from size changes. These include bifurcation systems (18) (19-21), and formal grammars (22-24).

The necessary implication would be that telencephalic computational operations are constrained to the set of those few mechanisms that do yield qualitatively different outcomes simply via quantitative change. We will show that there already are biologically grounded, bottom-up simulation and analytic studies that have strongly implicated formal grammars as the emergent mechanisms from telencephalic operation.

Again, we emphasize that it is not yet known whether some uniquely human cell types, or genetic innovations, or anatomical conformations, have the power to give rise to uniquely human abilities. It remains possible that new discoveries will identify mechanisms with the requisite power to explain novel human abilities. But we posit that it is not at all unreasonable to also search for explanations in the other direction: mechanisms by which new uniquely human abilities could arise purely from the rigid allometric increase in brain-body ratio and cortical-subcortical ratio. Rather than neglect this possibility, we take it as a serious contending hypothesis, and deliberately explore its potential implications.

## 2. Derivation of computations of thalamocortical and cortico-hippocampal circuits

Computational modeling of neural circuits has led to the identification of algorithms that may be carried out by particular anatomical systems in mammalian telencephalon.

Extensive bottom-up modeling work (14, 25-28) began with simulations of physiological operations occurring in the anatomical circuitry of thalamocortical loops. Every cortical region is connected to thalamic regions by both afferent and efferent projections (29-31);
Table 1 contains a brief précis of the simulated physiological steps occurring in thalamocortical circuitry reported by Rodriguez et al. (25):

**Table 1**
Simplified steps in simulated thalamocortical operation (from Ref [25]).

| | |
|---|---|
| i) | First input from periphery topographically activates core thalamic nucleus (Ct). |
| ii) | Ct topographically activates corresponding middle layers. |
| iii) | Activated middle layer modules vertically activate corresponding suprajacent layers, engaging lateral inhibition, producing statistical "clustering" response. |
| iv) | Output from superficial layers topographically activates deep layers. |
| v) | Diffuse feedback from L.V output to matrix thalamic nucleus (Mt). |
| vi) | Topographic feedback from L.VI output to thalamic nucleus reticularis and to Ct, selectively inhibiting the portion of the input corresponding to the "cluster" response. |
| vii) | Next input (or portion of input) arrives; Ct topographically activates middle, superficial, deep layers. |
| viii) | Layer V receives combination of non-topographic input from Mt, produced by prior input, and topographic activation from superficial layers produced by current input. |
| ix) | Intersection of these inputs selects sparse L.V response, and synaptic potentiation. |
| x) | Repeat steps v) to ix) until input completed. |

Many cortical areas receive only nonspecific (matrix), but not topographic (core), projections from thalamus; in these regions, outputs of superficial cortical layers become topographically-organized input to middle and superficial layers of downstream regions, whose deep layers send reciprocal feedback to the originating superficial layers (32-37). This cortico-cortical organization, together with the cortico-thalamic matrix loops that occur throughout cortex ((38), have been hypothesized to subserve corresponding functions (14, 39-43).

The resulting studies have strongly suggested characterizations of two constituent algorithms of thalamo-cortico-cortical operation: i) categorization of objects by similarity, and ii) chaining objects into sequences; pseudocode algorithms for these are presented in table 2 (from (25)):

The output of a given cortical area becomes input (both divergent and convergent) to other, downstream, regions, as well as receiving feedback from them. Producing categories and sequences, in cortico-cortical succession, yields sequences of categories, and categories of sequences of categories, etc. This hypothesized primary computation of thalamo-cortico-cortical circuitry (13, 14, 26, 44) is formally equivalent to grammars (45).

Extended thalamo-cortico-cortical networks produce successively nested sequences of categories (sequences of categories of sequences of categories), i.e., grammar expressions of increasing depth. This is concordant with findings that increasingly long auditory patterns are selectively processed by successively downstream cortical regions. A single category at any one cortical locus can itself be part of another entire sequence of categories.

**Table 2**
Derived simplified thalamocortical algorithms (from ref (25)).

```
Simplified thalamocortical core algorithm [from Rodriguez et al.]
  for input X
        for C ∈ win(X,W)
              Wj ⇐ Wj + k ( X - C )
        end_for
  X ⇐ X – mean(win(X,W))
  end_for
  where
        X = input activity pattern (vector); W = layer I synaptic weight matrix;
        C = responding superficial layer cells (col vector); k = learning rate;
        win(X,W) = column vector in W most responsive to X [∀j, max(X· Wj )]

Simplified thalamocortical matrix algorithm [from Rodriguez et al.]
  for input sequence X(L)
        for C ∈ TopographicSuperficialResponse(X(L))
              for V(s) ∈ C ∩ NNtResponse(X(L-1))
                        potentiate(V(s))
              NNt(L) ⇐ NontopographicDeepResponse(V)
              end_for
        end_for
  end_for
  where
        L = length of input sequence;
        C = columnar modules activated at step X(L);
        V(s) = synaptic vector of responding layer V cell;
        NNt(L) = response of nonspecific thalamic nucleus to layer V feedback
```

To illustrate the equivalence of sequences of categories on one hand, and simple grammars on the other, we may write down a simple grammar statement (typically written as a set of "rewrite rule" expressions):

$$X \to AB \mid A$$
$$A \to 5 \mid 23$$
$$B \to 2A$$

In this formalism, each left hand side is a category, which can be expanded to a sequence (possibly a sequence of length 1). X may be expanded to the sequence A followed by B; each member of that sequence is in turn a category (e.g., A), that may in turn be expanded to a sequence of categories (possibly a sequence of a single element, e.g., "5"), and so on. (The given grammar produces sequences that include 5, or 23, or 23223, or 525, …). Grammars are embedded categories of sequences of categories of sequences. The bottom-up simulation of biological thalamocortical mechanisms (25) directly led to two constituent algorithms whose interaction literally is a machine for constructing grammars.

It is notable that these formal grammars have no necessary relation to language. They are simply the mathematical constructs arising from simulated cortical-subcortical performance, operating on any inputs that occur, beginning with simple sensory inputs (e.g., visual, auditory), and continuing downstream via cortico-cortical successive regions, to construct ever-larger grammars, as the output of some regions fan in and fan out to become the inputs to further regions. As brains add ever more downstream cortical circuitry with evolutionary growth, the internal "representations" constructed by these mechanisms will grow larger. (It also is noteworthy that these mechanisms are generative, and yet can be less computationally costly than discriminative algorithms; see (44)).

Findings that successive cortical regions selectively process longer auditory patterns (46-48), as well as findings that specifically link hierarchical language syntactic structure to a successive hierarchy of cortical time scales corresponding to grammar construction (49), may all be seen as special cases of the general principle that successive cortical regions process incrementally expanded grammar rewrite-rule expressions. These expanded expressions will typically correspond to temporally longer auditory sequences, but exceptions to this correspondence may occur, in which case more-complex grammar expressions may correspond to briefer temporal sequences; the possibility of such divergent predictions may be empirically testable.

## 3. Parameters of biological grammar construction

What classes of grammars are being constructed by these cortical-subcortical mechanisms? Simple automata (e.g., finite state machines, or FSMs) produce simple languages (e.g., the set of "regular" grammars). Adding stacks to an automaton enables it to embed more complex patterns (e.g., $a^n b^n$, which includes strings (a b) and (a a b b) and (a a a b b b), etc.), yielding languages that are computationally more expressive, and correspondingly more powerful (e.g., context-free languages) (see Figure 2).

No cortical area operates in the absence of tightly coupled cortical-subcortical loops. In addition to ubiquitous cortico-cortical and cortico-thalamic loops, the cortex is enmeshed also in other prevalent cortical-subcortical loops, including cortical-striatal and cortical-limbic circuitry. Cortico-hippocampal loops are of particular interest in the present formulation. Measurement of hippocampal units during behavior has led repeatedly to findings in which the hippocampus produces response patterns consisting both of forward and backward "replay" and "preplay" of activations that previously occurred during behaviors (e.g., successive moves during navigation through a space)(50-55).

The potential utility of the "replay/preplay" hippocampal mechanisms has often been conjectured in the literature; we note here that this patterned operation may be consistent with what would be observed if sequential elements are being "pushed" onto, and subsequently "popped" back off of, a stack, enabling the tracking of ordinal positions of the elements within a sequence. This could be of adaptive utility especially for navigating paths through space, for memory "indexing," and for multiple other internal sequence-dependent functions (56-59).

If hippocampal operations compute stacks, cortico-hippocampal loops can be conjectured to be carrying out stack-using grammars (potentially including context free, tree-adjoining, or indexed grammars, as discussed in the next section). The way that stack mechanisms are used in grammars changes as a function of the "calls" to those stacks. In particular, a distinction is noted in the transition from context-free grammars, which use single stack mechanisms, to a family of grammars, called higher-order pushdown automata (HOPDAs), that is intermediate between context-free and context-sensitive grammars.

Turing machines compute the set of recursively enumerable grammars, and (modulo some debate in the literature, which will be discussed in the next section) they denote the top of the hierarchy of possible grammars, i.e., the most powerful of the family of automata. The computing power of Turing machines (and some other relatively powerful grammars) depends on the ability to read and write onto a highly flexible memory storage system; this typically takes the form of a "tape" of unlimited size. A more restricted memory storage mechanism is that of a pushdown stack: unlike a tape, the stack can only access the data that has most recently been placed onto it ("last in first out"), as opposed to the unrestricted reading from, and writing to, a Turing machine's tape. Grammars that use these more restricted memory systems can readily be shown to have lesser computational power than a Turing machine. A full Turing machine read-write tape can be constructed in several ways via specific modifications or additions to stack mechanisms, such as the presence of multiple independent stacks, which then are equivalent to the full unrestricted read/write ability of a tape (23, 24) (as opposed to single stack mechanisms).

HOPDAs fall short of the independent stacks that define more powerful grammars. They contain a more constrained form of multiple stacks: nested stacks, in which stacks are embedded dependently in each other, and cannot be used independently. HOPDAs thus form a natural class of grammars with more power than purely single-stack context free grammars and less power than grammars with multiple independent stacks such as context sensitive grammars (60).

Brains with comparatively small cortico-hippocampal ratios may use stacks, but only in an elementary way (e.g., in visibly-pushdown and context-free grammars); we hypothesize that larger cortico-hippocampal ratios enable stacks to be activated in a nested fashion, i.e., successive cortical "calls" first triggering a hippocampal stack, then holding its result cortically; then triggering another stack; holding that cortically; and so on. Moreover, higher cortico-hippocampal ratios may enable more cortical calls to the hypothetical hippocampal mechanism; the more calls, the more nesting of stacks may be possible. We generalize these hypotheses to conjecture that increased cortico-hippocampal ratios may correspond to increased nesting of stacks. The smallest cortico-hippocampal ratios may be insufficient to enable nested stacks at all. Slightly larger cortico-hippocampal ratios may suffice to enable nested stacks, albeit possibly very few. Larger ratios may enable larger numbers of nested stacks. As will be seen, this sequence of quantitative changes may be associated with qualitatively different families of grammars from context free, to tree-adjoining, to indexed grammars.

As mentioned, the cortical-subcortical grammars educed here are applicable to sensorimotor (and other) data, with no required relation to language. Grammars are, of course, well known to characterize the syntactic aspect of the human language faculty (61-64). The strong evidence of absence of human-level syntactic language capabilities in other animals (and the apparently abrupt or "saltatory" manner in which the new capability arises in the evolutionary transition to humans), has been quite rationally used to strongly suggest that there must be some new faculty arising in humans that engenders the new species-specific competence, yet as noted the evidence from comparative anatomy instead indicates strong allometric constraints. There may of course be language-specific exceptions to these allometries, yet extensive searches for human anatomical specializations (65-74) still show no mechanisms that in any way hint how they could yield the striking characteristics of human language.

We began by introducing two distinct mechanisms, hierarchical categorization and sequence chaining, each arising from biologically-based simulations of cortical-subcortical operation, independent of functional or behavioral considerations, which give rise to a unified mechanism – grammars – for vision, for action, and for other cognitive functions (see discussions in (14, 25, 26)). The described modeling work did not start with behavioral data nor with any formal theoretical stance. Rather, from bottom-up biological modeling of thalamo-cortical and cortico-cortical circuitry, the two constituent algorithms of hierarchical categorization and chaining were observed (25) and subsequently studied in simplified form (75), directly giving rise to algorithms in the family of grammars (14, 26, 44). Although different cortical regions are often "assigned" differential responsibility for particular behavioral and cognitive functions (based in part, for instance, on lesions or imaging studies), it has repeatedly been hypothesized that these different circuits are constructing the same computational entities, with their differential responses arising solely from connectivity. Starting from purely perceptual information such as visual and auditory data, the operation of different pathways through the same kinds of thalamocortical and cortico-cortical circuits may be building up successively larger hierarchical structures, leading to representations of complex entities such as faces, places, houses, animals, in different path locations throughout the brain. Proceeding through further brain regions, they may continue to construct ever larger grammar-based relations among memories. Indeed, the remarkable consistency of anatomical architectures throughout neocortex has often led to questions of how such closely consistent circuitry could lead to such a wide variety of different functions. A range of hypotheses, although differing in many particulars, may be seen to share many of these fundamental points of agreement (41, 75-77).

## 4. Characterization of different capabilities of different grammars

Altering a given grammar via seemingly modest add-ons (different stack systems) yields mechanisms with irreducibly altered new capabilities (61, 78, 79). Simple finite state machines (FSMs) are capable of recognizing the family of regular grammars, a set of relatively simple mathematical forms (45, 80). Simply adding a stack memory to FSMs produces "pushdown automata," (PDAs), which recognize context-free grammars, a larger family that cannot be reduced to regular grammars (22-24).

The progression continues: incorporating additional nested stack mechanisms into FSMs yields successively higher-order PDAs, (HOPDAs), from level-1 HOPDAs with a single nested stack (which recognize grammars larger than context-free), through level-n HOPDAs with up to an infinite number of nested stacks (24, 81, 82), which recognize the still-larger set of "indexed" grammars (60, 83-88) (Fig 2).

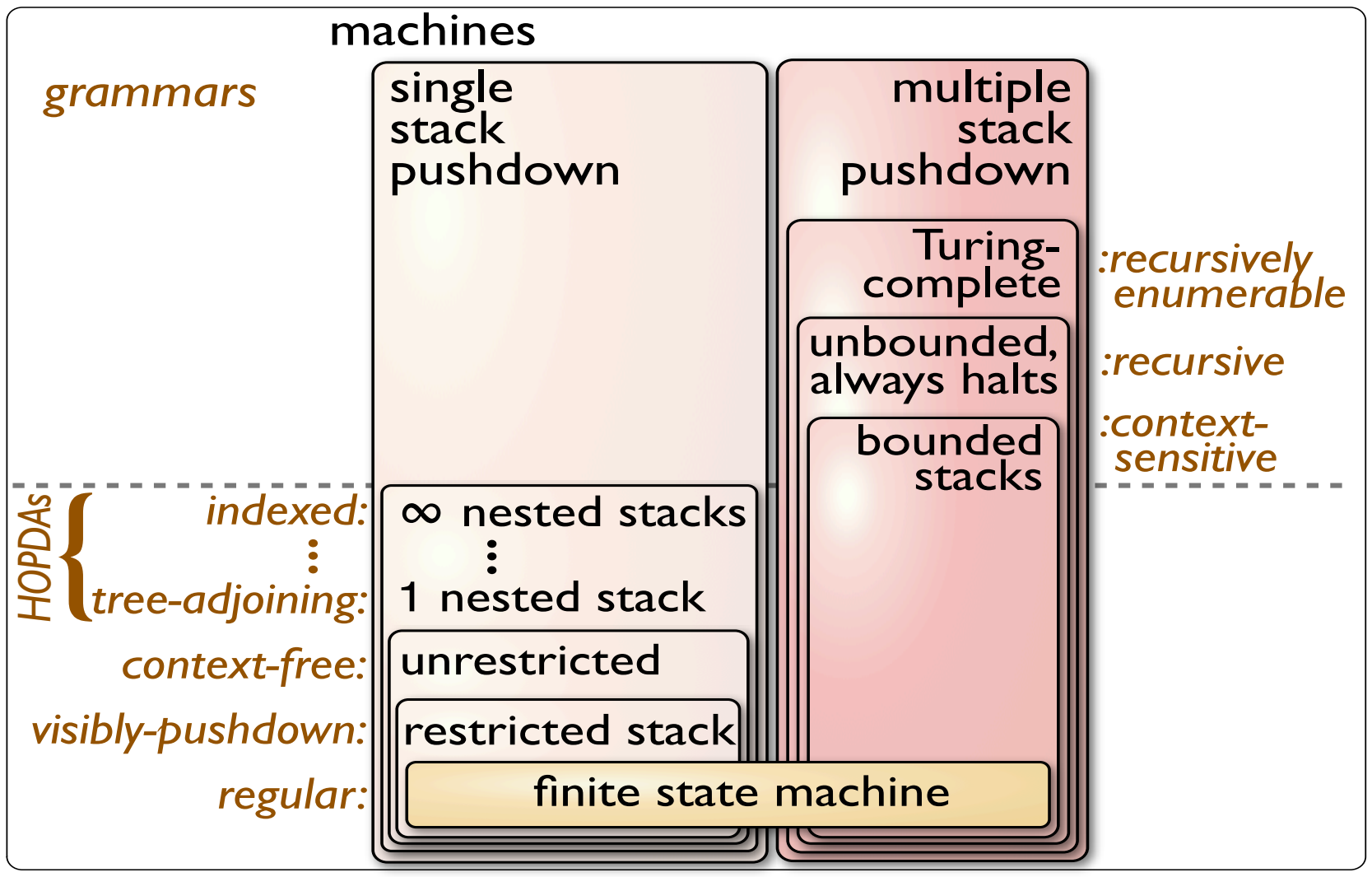

**Figure 2.** Hierarchies of machines (automata) and the grammars they recognize. Finite state machines (FSM) recognize "regular" grammars. (Left) Adding a single stack pushdown memory to an FSM produces systems that recognize visibly-pushdown and context-free grammars (for restricted and unrestricted stacks, respectively); adding a single "nested" stack (see text) produces higher-order pushdown automata (HOPDA); added nested stacks produce machines that can recognize successively larger grammars. (Right) A distinct family of still more powerful automata can be constructed by adding multiple independent stacks (which need not be nested), creating another hierarchy of automata that recognize context-sensitive grammars, recursive grammars, and recursively enumerable grammars.

Whereas all intuitively "effective calculations" (in an informal sense) are able to be computed by a Turing machine, it is not trivial to discover whether some given operation may require the full power of a Turing machine, or only require a proper subset of such machines, such as linear bounded automata. The latter (shown as "bounded stack" machines on the right side of Figure 2) compute context sensitive grammars, which are a subset of full recursively enumerable grammars (i.e., those requiring a Turing machine).

Several well-studied grammars that are more powerful than indexed grammars can be constructed, but not by adding further nested stacks. Instead, adding multiple, separate, independent stacks can produce systems that recognize languages larger than those of indexed grammars, up through recursively enumerable languages, recognized by Turing machines (Figure 2, right side). This distinct mechanism of multiple stack systems, as opposed to single-nested-stack systems, forms a separate clade of the mathematical family of grammars, setting apart those on the left of Figure 2 from those on the right. A relevant question is: do human brains (or other mammalian brains) exhibit the power of grammars such as context-sensitive grammars, or recursively-enumerable grammars, that are larger than indexed grammars?

Many specific instances of nested-stack HOPDAs have been proposed in the scientific literature to meet the criteria of human natural language processing (which requires grammatical capacities beyond that of a context-free grammar; (89, 90) ): the four most-studied suggestions in that literature are tree-adjoining grammars (91), head grammars (92), combinatory categorial grammars (93, 94), and linear indexed grammars (a proper subset of full indexed grammars; (95)). It is compelling to note that these have all been shown to be nested-stack higher-order PDAs (HOPDAs) (81, 82, 96).

It is important to note the distinction between a mechanism, on one hand, and the capacity of the representations that can be constructed by that mechanism on the other. The computational power of any grammar of a given class (e.g., tree-adjoining) is specified without reference to the physical instantiation of a mechanism that may implement it. Different mechanisms may both compute the full set of recursively enumerable grammars, with one doing so in an excruciatingly laborious manner while the other does so efficiently; any such distinctions between them are completely separate from their designation as Turing complete. For instance, most programming languages are Turing complete yet can be very difficult to apply successfully to many everyday tasks; as computer scientist Alan Perlis admonished us: "Beware the Turing tarpit, where everything is possible and nothing is easy" (97). Ongoing work is aimed at further studies of biological mechanisms potentially giving rise to the grammars described here (98). It is notable that a number of researchers have studied the formal grammar capabilities of a range of entities such as differently-configured artificial neural networks (ANNs) and other hypothetical neural assemblies (see, e.g., (99-101)).

## 5. Specific implications of grammatical growth

If mammalian cortico-cortical and cortical-subcortical circuits carry out the processing of simple grammars (14, 26), we may forward a specific chain of hypotheses concerning telencephalic size.

First, it is possible that the computational power of mammalian precursors (including reptiles and perhaps the therapsid cynodonts) is characterized by finite state machines with no stack memory, thus perhaps producing at most regular grammars (Figure 2).

When mammals diverged from the reptiles some 200 million years ago, two entirely new systems appeared: neocortex (embedded in thalamo-cortical loops), and the full hippocampus (beyond the CA3 of reptiles to incorporate the full mammalian loop of entorhinal-CA3-CA1-subiculum-entorhinal (102-104)).

We have described bottom-up simulations of thalamo-cortico-cortical loops directly producing simple grammars, in the form of hierarchically organized sequences of categories, whereas cortico-hippocampal loops have been described as exhibiting play–replay capabilities that we have suggested may be indicative of stack mechanisms that can be played forward (push) and backward (pop).

From the evolutionary inauguration of these mechanisms, going forward, the allometric increase of cortical to subcortical ratio that occurs with increasing mammalian brain size (1, 6, 7) (Figure 1) adds disproportionate brain resources to cortical-subcortical circuits, which may correspond to increased capacity for nesting of stacks, i.e., higher cortical-subcortical ratios increase the computational power of a brain, potentially producing nested-stack level-n HOPDAs with successively larger values of n.

We might have begun from a more top-down point of view by positing that humans must have Turing-level capabilities, and that, perhaps, added cortical processing adds to a Turing tape-like or multiple-independent-stack mechanism in brains. Indeed, we might well have assumed that brains could readily contain multiple independent stacks, or any other type of read/write memory system, if our hypotheses had been based solely on generalized notions of the distributed nature of brain networks, unconstrained by measures of behavioral or computational capabilities, or by detailed modeling of brain architectures and operations.

Yet the extensive studies of the computational power of human language conclude that our most powerful innate capability – natural language – does not exceed that of HOPDAs – and thus requires neither tapes nor independent stacks. Multiple independent stacks, although they might have been expected to occur in the parallel distributed networks of the brain, are unnecessary for any of the analyses forwarded here, based either on brain circuit analysis or on observed and measured behaviors. Accounting for any known innate human behaviors requires only single or nested pushdown stacks, but not multiple stacks (nor read/write tape mechanisms). If all mammals with smaller brain-to-body ratios than humans exhibit the abilities solely of single or (at most) nested pushdown stacks, then the known constraints of allometry and comparative anatomy strongly suggest that humans are similarly constructed.

In fact, if multiple stacks had somehow been introduced for the first time in mammalian evolution, then some allometric divergence, or some anatomical anomaly, might have been expected to accompany such a new system; yet no new mechanisms of this kind have been identified. Although a wholly-new mechanism such as multiple independent stacks could have occurred, the biological modeling that we have described finds no evidence-based reason to introduce these new mechanisms, nor does any behavioral or cognitive measure, including full human natural language.

It is often posited that humans exhibit full recursively-enumerable, Turing-complete capacities, which would require the equivalent of (at least) unbounded multiple-stack mechanisms to be present in some form. (As noted, these are mathematical equivalences; there need not be any such thing as a specific implementation of e.g., independent stacks, but, rather, some mechanism that is of equivalent power.) Indeed, it has even been hypothesized that human abilities may exceed those of a Turing machine (105-108), though there is a specific question of whether these models require infinite precision computations: i.e., perhaps non-Turing computable outputs are attainable only by adopting the confounding assumption that non-Turing-computable inputs are somehow permitted (109).

But it is noteworthy that actual quantification of human behavioral abilities (or those of other organisms) has been notoriously elusive. The behavioral repertoires of even closely related species can often be qualitatively distinguished from each other. Work on quantification of behavioral and cognitive capabilities exhibits promise for relating these measures to brain characteristics (see, e.g., (110-113), though as yet, do not know where we might locate them in the ranking of abilities characterized by the grammar hierarchy (nor other candidate methods of quantifying mental capacity).

Multiple experiments have tested the abilities of particular species to acquire sequence information containing particular structure (such as birds being shown to learn sequences constructed with particular grammatical patterns, e.g., (114-117).

But the ability to forcibly learn specific subsets of artificial grammars is distinct from having the capacity for full grammars. In general, we wish to be careful to avoid the category error of observing a subject learning patterns of a particular complexity and then concluding that the subject must therefore be capable of learning all and only patterns of that complexity.

In particular, there are many instances in which critical controls have not been run. For instance, showing that a subject learns a set consistent with $a^n b^n$ has sometimes been used to argue that that subject is therefore capable of learning at least context-free languages, but this data is insufficient: crucially, it must also be shown that the subject would at the same time *not* accept a member of $a^n b^m$ . In the absence of this control, there are alternative mechanisms with lower computational power that could explain the same results (see, e.g., (118, 119)).

In general, when trying to infer computational grammar capabilities, it is important to note that experimental evidence can, at best, weaken a hypothesis, but is insufficient to fully verify (let alone discover) the existence of particular computational power (see, e.g., (120)). As a trivial example any complex grammar can be simulated perfectly, up to a fixed embedding depth, by a weaker grammar. For instance, an indexed grammar of any given *fixed* size, can be straightforwardly simulated by a sufficiently large finite state machine (consisting of no stacks). Thus the field necessarily distinguishes between capacity that can be directly demonstrated vs. capacity that can be inferred. Errors of inference can readily occur in both directions: any single empirical demonstration may seem to imply a higher capability than can actually be logically justified, or such a demonstration may appear to imply a simpler mechanism than would be justified based on consideration of additional observations.

As mentioned, most behavioral and cognitive abilities are notoriously difficult to quantify with any accuracy; most measures of behavior or cognition are still underspecified, and may be unknowable at least with any current measurement tools. One outstanding exception is natural language. Multiple researchers from disparate starting points have repeatedly converged on the family of HOPDAs: nested-single-stack pushdown grammars (81, 90-92, 94-96, 121), of which several distinct formulations have been shown to be formally equivalent (96). And yet the generative expressive power of human natural language is one of the few remaining capabilities that has been consistently shown to exceed that of other mammals. This naturally raises the question of just what computational powers are being referred to when humans are posited to have the power of Turing machines?

We specifically conjecture that the innate computational power of human brains is equivalent not to Turing-complete (recursively enumerable) systems, but rather to indexed grammars at most, and, quite possibly, even smaller mechanisms.

At first glance, this assertion seems plainly false. Humans construct digital computers, which are capable of acting as Turing-equivalent, so surely humans are – at least! – themselves fully Turing equivalent. Indeed, as mentioned, conjectures have been forwarded that human brains may be of super-Turing computational power, e.g., (105-107). Yet actual occurrences of instantiated Turing-equivalent systems are remarkably sparse. In fact, even instances of machines that accept any more than full context-sensitive languages (far smaller than recursively enumerable languages) are exceedingly rare. An everyday example: typical computer programming languages have the full power to express recursively enumerable grammars, yet any computer (e.g., a PC) is actually equivalent to a linear bounded automaton, capable only of accepting the set of context-sensitive grammars. The addition of infinite storage would convert a computer to full Turing completeness. Such distinctions may appear picayune, but consideration of detail is the crucial distinguishing factor in this instance. The reason that this theoretic limit on computers is not noticeable in everyday usage is that it has little to no pragmatic effect: any computer can perform any specified finite set of Turing-equivalent computations, though it may not do so within the lifetime of the human race (or the universe). Turing completeness radically exceeds the requirements of almost any everyday specifiable task.

In fact, it is difficult to pose any pragmatic, practicable task that requires any mechanism more powerful than indexed grammars (91). The less-powerful grammars, e.g., context-free grammars, contain some relatively glaring limitations, which are surpassed by the nested-stack family (the HOPDAs, or higher order pushdown automata). Indeed, these higher order automata are sufficiently powerful that even the weakest of them (level-1 HOPDAs, or tree-adjoining grammars) have been argued to suffice for full natural language processing (81). It is easy to overlook the extraordinary power of sub-Turing mechanisms.

Yet, as mentioned, humans do of course produce Turing-complete things like programming languages. In such cases, do we overcome the limitations of HOPDAs, and if so, how is that possible? We have already pointed out the jump in power that comes with access to a potentially infinite memory store (a tape, or multiple independent stacks). When humans do logic, as when we solve advanced mathematical problems, or build programming languages, or construct complex engineering systems, we can use external aids, and thus may exceed our innate capacities. A Turing machine is nothing more than a simple finite state machine, with the addition of a potentially infinite memory store. And humans with paper and pencil constitute a system far more powerful than humans with nothing but their innate systems. This observation is (perhaps uncomfortably) suggestive: we may think of language as the defining extra feature setting humans apart from other animals, yet the major human accomplishments that have actually set us apart are dependent on more than just language-level abilities: constructing buildings, cars, aircraft; organizing into lawful polities; discovering and mastering laws of nature; all are feats that are highly unusual for any typical human, and most of them necessitated extensive cognitive work beyond just language, requiring external aids, either from scientific instruments, computers, or at least paper, pencil, and extremely abnormal attention to detail. Many of the notable accomplishments of the human race simply could not be replicated by most members of the human race: building a house, a car, an office, a computer, even a pencil, are beyond the capacity of almost anyone. Humans may innately be capable of the impressive and unique power of generative language. Much more rarely, with external aids, some humans have been capable of far more.

Several consequences are entailed by these analyses. We briefly highlight four of these:

Successive cortical areas process increasingly complex grammars. Outputs of a given cortical region are inputs to successively downstream regions; early regions will be selective to relatively simple features whereas later regions will process increasingly longer and more complex grammar formulations; this is consistent with findings of the selective sensitivity of cortical regions to longer and more complex auditory patterns in language and in music (46-49).

Brains must exceed a specific size threshold to attain human language. Mammalian brains of sufficient size will achieve the computational power of HOPDAs. These will be adequate for processing the syntactic structures of human language, whereas below that size, the mechanisms may suffice for recognition of simpler structures but not for human syntax.

Size of information vs. size of processing mechanism. A given brain has the computational power to process data with complexity up to, but not exceeding, that brain's grammar class capacity. If presented with less-demanding data, i.e., data structures that do not require the full power of that brain's grammar class, then representations will be constructed that correspond to the complexity of the data, despite that brain's excess computational capacity. Thus, for instance, human syntactic data will cause a sufficiently-powerful system to construct high-order PDAs (such as tree-adjoining or combinatory categorial grammars) whereas, e.g., phonological data, presented to the same system will yield at most regular grammars, because that is all that the phonological data require. This hypothesis contrasts with those suggesting that data of different complexities must necessarily be processed by different mechanisms (122). We specifically hypothesize that phonological data is processed by the same mechanisms as syntactic data – the difference being solely that phonology data only yields construction of FSM complexity to be fully processed, so no larger grammatical structures will be learned by the system in response to that data (also see (123)). The system is driven by its input data to construct models of that data, up to the limit of the grammar class of the particular brain mechanism.

Diminishing returns of increasing brain size. Once species' brain sizes have reached the ability to produce sufficiently high order grammars (HOPDAs), further additions to its cortical-subcortical ratio will asymptotically approach full indexed grammars, incrementally conferring less computational power per added nested stack. If, as we have hypothesized, mammalian brains indeed contain single stack rather than multiple stack pushdowns, then there is a point beyond which brain size increases will yield comparatively little added computational power. This may contribute to an explanation of why hominin brains increased for roughly four million years but then peaked and retreated just within the last twenty to thirty thousand years (13, 14, 124-126).

As mentioned, grammars characterize the nature of the family of representational constructs that can be built, without specifying *how* they are built, nor how efficiently they may be implemented. In particular, depending on the representational codes that may be constructed within a brain, it may be possible for some tasks to be more rapidly carried out in one brain versus another brain, despite both being potentially characterizable by particular equivalent HOPDA grammars. In addition, allometric ratio differences observed in different orders such as rodents versus primates (7) may indicate architectural differences within the overall mammalian brain plan. Nonetheless, the striking allometric regularities of mammalian brains strongly suggest that identical or closely-related computational means are employed across multiple mammalian orders.

## 6. So, are there brain specializations for human language?

Theorists formerly proposed multiple abilities that might be unique to humans: tool use, war, culture, reasoning, theory of mind, cooperation – all of these and many more have since been shown to occur in non-human species. Oddly, the one capacity that appears still to powerfully resist identification in other species is that of human language: though many animals have elaborate communication systems, a large body of evidence suggests that only humans can generatively manipulate arbitrarily complex concepts via a fixed vocabulary, by virtue of the syntactic structure of human language (127, 128). Due to the well-characterized distinction between the syntactic structure of human languages versus the structures of other animals' communication systems (61-63, 79) it has long been posited that there must be specialized, unique brain circuitry endowing humans with our singular linguistic capabilities (64, 129-131).

If brains mosaically acquire domain-specific specializations, then distinct tasks may be carried out by distinct brain mechanisms, whereas if our capabilities mostly arise from concerted brain size increases, then even tasks that appear very different may be carried out by shared mechanisms. The question is especially relevant to tasks of demonstrably different complexity: must easier versus harder computations be accomplished by different means? This can be posed as both a within-subject and a between-subject question: i) within an organism, are tasks of different complexity executed by distinct mechanisms of corresponding complexity; and ii) if different organisms exhibit behaviors of distinct complexities, must they be using different brain mechanisms to do so?

The search for candidate brain specializations for language has to date identified slight variations in certain cell types (65, 66), circuit architectures (10, 67-69, 71, 73, 74, 132), and gene alleles (70, 72, 133-136);

none of these can yet suggest any actual mechanisms by which these seemingly exiguous differences could generate uniquely human language abilities. (We note that the brain computations being discussed are not in any way tied just to the processing of external stimuli, i.e., so-called purely "reflexive" processing (137), but incorporate all cortico-thalamic and cortico-cortical internal operations, of which the "reflexive" stimulus-processing operations are just a small minority. The present analysis arises from these telencephalic models, and nothing in the analysis is intended to limit them to just sensory processing.)

Evidence for specialized localization in the human brain exists for many capabilities including language, and face recognition. Yet the existence of such localities does not imply evidence of their innateness. For instance, a focal occipito-temporal cortical region (the VWFA, or visual word-form area) exhibits strongly preferential responses to the presentation of written words. It would provide ready evidence of the innateness of reading – if we did not know that reading is culturally inherited, requires substantial training (unlike auditory language), only arose within the last several thousand years, and that learning reading, even in adulthood, creates profound changes in cortical responses (138-140).

Human syntactic systems require at least level-1 HOPDA mechanisms to process. By contrast, it is known that phonological systems only require far weaker mechanisms (122). This difference characterizes the data that is input to the mechanism: phonological data can be parsed with weaker grammar mechanisms than human syntactic structures can. Because these data present different challenges to processor, it has been argued that different mechanisms must be at play in processing them. This is not a necessary conclusion. If a too-powerful mechanism is presented with simple data, it still is capable of identifying structures consistent with that simple data. (Perhaps analogously, a fixed machine learning algorithm such as an SVM or CSL method can readily identify either simple patterns (in simply-organized data), or complex patterns (in data containing complex organizations); as long as the mechanism is capable of identifying sufficiently complex patterns, it can recognize those or any simpler patterns). Thus it is possible that the same (sufficiently-powerful) grammar mechanisms may be brought to bear against phonological and syntactic data, learning simpler structures for the former and more complex structures for the latter.

## 7. Summary and predictions

In sum, we specifically conjecture that

- repeated cortical-subcortical loops carry out the equivalent of a single-stack pushdown automata (PDA) mechanism throughout much of mammalian telencephalon;
- successively downstream cortical regions correspond to longer grammars;
- the documented allometric increase in cortical-subcortical ratio for bigger brains can increase their computational capacity to the level of higher-order grammars;
- evidence suggests that humans are the only organisms that may reach the equivalent of HOPDAs (nested stack grammars);
- evidence does not indicate that humans or other organisms achieve context-sensitive or higher grammars.

There are potentially predictive consequences for non-human animals. If non-humans' capacities are context-free or less, their limitations may be revealed via analysis of their vocalization or learning capabilities (141). Existing empirical evidence is consistent with these sub-context-free limits for nonhuman primates (129) and for birds (117, 141-143). (It can be noted that "compositional syntax" is indeterminate with respect to the grammar hierarchy; the regular grammars of simple finite state machines can exhibit compositionality.) We again emphasize that empirical evidence can, at best, weaken a hypothesis of representational capacity, but is always insufficient to fully verify it (120).

Intriguingly, evidence exists that the level of context free grammars may be attained in whale song (144). Though many differences exist between primate and cetacean cortices (145-147), nonetheless cetaceans' brains follow the mammalian pattern, and cetacean brain-to-body ratios are second only to humans (5), exceeding those of all nonhuman primates. This represents an intriguing unaddressed question in comparative capabilities, which may be a fruitful topic for further study.

Throughout these analyses, we consistently find evidence consistent with the hypothesis that small mammalian brains (with smaller cortical-subcortical ratios) may be capable only of tasks with limited computational requirements whereas big brains (with larger cortical-subcortical ratios) may be able to perform wholly new tasks, despite using the same fundamental mechanisms as smaller brains – just more of them.

The hypotheses outlined here offer a specific explanatory account of apparently abrupt (saltatory) changes to behavioral and cognitive capabilities as brains evolved: quantitative increase of stacks in single-stack pushdown grammars do yield qualitatively more powerful computational mechanisms. We forward these hypotheses as a unifying formal account of i) how repeated similar brain algorithms can be successfully applied to apparently-dissimilar computational tasks (e.g., perceptual versus cognitive; phonological versus syntactic); and ii) how quantitative increases to brains can confer qualitative changes to their computational repertoire.

**Acknowledgments.** This work was supported in part by grants from the Office of Naval Research and the Defense Advanced Research Projects Agency. We thank Hava Siegelmann, and Selmer Bringsjord, both of whom provided very useful discussions (though neither necessarily agrees with us).